\documentclass[prb,aps,twocolumn,showpacs,floatfix]{revtex4-1}
\usepackage{epsfig}
\usepackage{amsmath}
\usepackage{mathptmx}
\usepackage{eucal}
\usepackage{bm}
\usepackage{color}
\usepackage{epstopdf}
\usepackage{graphicx}
\usepackage{natbib}

\begin{document}

\title{Spin-polarized Josephson and quasiparticle currents in
superconducting spin-filter tunnel junctions}
\author{F. S. Bergeret$^{1,2}$, A. Verso$^{2}$ and A. F. Volkov$^{3,2}$ }

\affiliation{Centro de F\'{i}sica de Materiales (CFM-MPC), Centro Mixto CSIC-UPV/EHU, Manuel de Lardizabal 4, E-20018 San
Sebasti\'{a}n, Spain\\
$^2$Donostia International Physics Center (DIPC), Manuel
de Lardizabal 5, E-20018 San Sebasti\'{a}n, Spain\\
$^3$Theoretische Physik III, Ruhr-Universit\"{a}t Bochum, D-44780 Bochum, Germany\\}

\date{\today }

\begin{abstract}
We present a theoretical study of the effect of spin-filtering on the
Josephson and dissipative quasiparticle currents in a superconducting tunnel
junction. By combining the quasiclassical Green's functions and the
tunneling Hamiltonian method we describe the transport properties of a
generic junction consisting of two superconducting leads with an effective
exchange field $\mathbf{h}$ separated by a spin-filter insulating barrier.
We show that besides the tunneling of Cooper pairs with total
spin-projection $S_{z}=0$ there is another contribution to the Josephson
current due to {triplet Cooper pairs with total spin-projection} $%
S_{z}\neq 0$. The latter is finite and not affected by the spin-filter
effect provided that the fields $\mathbf{h}$ and the magnetization of the
barrier are non-collinear . We also determine the quasiparticle current for
a symmetric junction and show that the differential conductance may exhibit
peaks at different values of the voltage depending on the polarization of
the spin-filter, and the relative angle between the exchange fields and the
magnetization of the barrier. Our findings provide a plausible explanation
for existing experiments on Josephson junctions with magnetic barriers,
predict new effects and show how spin-polarized supercurrents in hybrid
structures can be created.
\end{abstract}

\maketitle


\affiliation{Centro de F\'{i}sica de Materiales (CFM-MPC), Centro
Mixto CSIC-UPV/EHU, Manuel de Lardizabal 4, E-20018 San
Sebasti\'{a}n, Spain\\
$^2$Donostia International Physics Center (DIPC), Manuel
de Lardizabal 5, E-20018 San Sebasti\'{a}n, Spain\\
$^3$Theoretische Physik III, Ruhr-Universit\"{a}t Bochum, D-44780 Bochum, Germany\\
}

\bigskip

\textit{Introduction} The prediction of long-range triplet superconducting
correlations in superconductor-ferromagnet (S-F) hybrid structures\cite%
{Bergeret:2001bk,Bergeret:2005bf} has led to intense experimental activity
in the last years \cite%
{Keizer:2006jw,Khaire:2010ea,Robinson:2010gi,Klose:2012bf,Anwar:2010bs}.
These experiments have shown that a finite Josephson current can flow
between two superconductors connected by a ferromagnetic layer whose
thickness far exceeds the expected penetration length of singlet pairs. The
Josephson current measured in these experiments is attributed to the flow of
Cooper pairs in a triplet state. According to 
the theory, the appearance of
triplet correlations occurs only in the presence of a magnetic inhomogeneity
located in the vicinity of the SF interface \cite%
{Bergeret:2001bk,Volkov:2003hl,Eschrig03,Houzet:2007ib,Footnote}.
The inhomogeneity can be either artificially created \cite{Khaire:2010ea} or
can be an intrinsic property of the material, as for example the domain
structure of usual ferromagnets \cite{Klose:2012bf} or the spiral-like
magnetization in certain rare-earth metals \cite{Robinson:2010gi,Sosnin2006}.

The Josephson triplet current is nothing but a dissipationless
spin-polarized current and therefore its control would be of great advantage
in the field of spintronics \cite{Eschrig:2011ht}. Important building blocks
of spintronic circuits are magnetic insulating barriers with spin-dependent
transmission, so called spin-filters (I$_{sf}$), which have been studied in
several experiments using for example europium chalcogenide tunnel barriers%
\cite{Tedrow:1986hy,Moodera:1988bb,Hao:1990fl,Santos:2008cm} The question
naturally arises whether one can use these spin-filter tunneling junctions
to control and eventually to create a triplet Josephson current. We will
address this question in the present letter.

In spite of several studies of the transport properties of spin-filter
tunneling barriers, the Josephson effect has only recently been explored
through a S-I$_{sf}$-S \cite{Senapati:2011fm}. The tunnel barrier used was a
GdN film that reduced the value of critical current $I_{c}$ compared to a
non-magnetic barrier. Beside the large reduction of $I_{c}$ the authors of
Ref. \cite{Senapati:2011fm} also observed that the $I_{c}(T)$ curve deviates
at low temperature from the expected tunneling behavior\cite%
{Ambegaokar:1963fk}. Teoretically, the effect of spin-dependent transmission
on the Josephson current was first considered by Kulik \cite{Kulik:1966us}
and Bulaevskii et al \cite{bula_kuzi} on the basis of the tunneling
Hamiltonian. It was demonstrated that spin-selective tunneling always leads
to a reduction of the critical current with respect to its value in the spin
independent case {or even to the change of sign of the critical current}.
Later on it was shown that the magnetic barrier in a I$_{sf}$-S structure
induces an effective exchange field in the superconductor \cite%
{Tokuyasu:1988et, Cottet2009}. Other theoretical works have addressed the
Josephson effect through spin-active barriers in ballistic systems \cite%
{Fogelstrom:2000ek,Cottet:2005hk,Kalenkov:2009ih} and through ideally
ballistic superconductor-ferromagnetic insulator-superconductor junctions %
\cite{Tanaka:1997dx,Kawabata:2010im}. Also the spin-polarized current
through S-N-F junction has been studied in Refs.\cite%
{HuertasHernando:2002gj,Giazotto:2008ju}. However, none of these works
presented a {comprehensive }theoretical study of the Josephson effect
by taking into account both the spin-filter effect and the presence of the
exchange field in the superconducting electrodes. Nor has the interplay
between spin-filtering and triplet supercurrents been investigated.

The aim of the present letter is to provide a complete description of the
transport properties of Josephson junctions with spin-filters. For that sake
we introduce a simple model which allows us on the one hand to derive simple
and useful expressions for the dc Josephson and quasiparticle currents in a
S-I$_{sf}$-S and on the other hand to predict the conditions under which the
creation of a spin-polarized super-current is possible. Our model considers
the spin-filter effect of the I$_{sf}$ barrier and a finite exchange field
in the electrodes.  We show that the contribution to the current from Cooper
pairs in the singlet and triplet states with zero spin projection vanishes
in the case of a fully spin-polarized barrier.  However, the contribution to
the Josephson current from tunneling of Cooper pairs in a triplet state with
non-vanishing spin projection is independent of the strength of the
spin-filter.  The latter contribution is finite provided that the exchange
fields in the electrodes and the spin quantization axis of the barrier are
non-collinear.  This remarkable result explains how spin-polarized currents
can be created and controlled by means of spin-filter barriers.  We also
calculate the differential tunneling conductance of the S-I$_{sf}$-S
junction, and analyze how the Zeeman-splitting peaks depend on both the
spin-filter parameter and the exchange field in the electrodes. Our model
allows a quantitative description of existing transport experiments on S-I$%
_{sf}$-S junctions \cite{Tedrow:1986hy,Moodera:1988bb,Hao:1990fl} and gives
a possible explanation for the temperature dependence of the critical
Josephson current observed in Ref. \cite{Senapati:2011fm}, We finally
discuss the applicability of our model to real systems.

\textit{The Model} We consider a tunnel junction between two superconductors
(see inset of Fig. 1). The tunneling barrier, black area in Fig. 1, is a
spin-filter.  The grey regions close to the barrier are thin ferromagnetic
layers with a finite exchange field acting on the spin of the conducting
electrons. The direction of these fields is arbitrary. We assume for
simplicity that the thickness of the superconductors is smaller than the
coherence length. In this case one can average the equations for the Green
functions over the thickness and get a uniform superconductor with built-in
exchange field\cite{Bergeret:2001ba}. Under these assumptions the system is
described by a generic Hamiltonian  which is homogeneous in space:
\begin{equation}
H=H_{R}+H_{L}+H_{T}  \label{Hamiltonian}
\end{equation}%
where $H_{R(L)}$ describes the left and right electrodes consisting of a BCS
superconductor with an intrinsic exchange field. For example for the left
electrode it reads
\begin{equation}
H_{L}=\sum_{k,s,s^{\prime}}a_{ks}^{\dagger }\left[ \xi
_{k}\delta_{ss^{\prime}}- (h_{L}\mathbf{n_{L}.\hat{\sigma}})_{ss^{\prime }}%
\right] a_{ks}+\sum_{k}\left( \Delta _{L}a_{k\uparrow }^{\dagger
}a_{-k\downarrow }^{\dagger }+h.c.\right) \; ,
\end{equation}%
where $a(a^{\dagger })$ are the annihilation (creation) operator of a
particle with momentum $k$ and spin $s$, $\xi _{k}$ is the quasiparticle
energy, $\Delta $ is the superconducting gap, $\mathbf{\sigma }=(\sigma
_{1},\sigma _{2},\sigma _{3})$ is a vector with the Pauli matrices, $h_{L}$
the amplitude of the effective exchange field and $\mathbf{n}$ a unit vector
pointing in its direction. The $H_{T}$ term in Eq. (\ref{Hamiltonian})
describes the spin-selective tunneling through the spin-filter and is given
by
\begin{equation}
H_{T}=\sum_{s,s^{\prime }}\left( \mathcal{T}\mathbf{\hat{\sigma}}_{0}+%
\mathcal{U}\mathbf{\hat{\sigma}}_{z}\right) _{ss^{\prime }}a_{s}^{\dagger
}b_{s^{\prime }}+h.c.
\end{equation}%
where $a$ and $b$ are the field operators in the left and right electrodes
respectively. $\mathcal{T}$ and $\mathcal{U}$ are the spin independent and
spin dependent tunneling matrix elements. We neglect their momentum
dependence. The tunneling amplitude for spin up (down) is then given by $%
T_{\uparrow (\downarrow )}=\mathcal{T}\pm \mathcal{U}$. We assume that the
origin of the different tunneling amplitudes is the conduction-band
splitting in the ferromagnetic insulating barrier which leads to different
tunnel barrier heights for spin-up and spin-down electrons \cite%
{Hao:1990fl,Santos:2008cm}.

In order to calculate the current through the junction it is convenient to
introduce the quasiclassical Green functions $\check{g}_{R(L)}$ for the left
and right electrodes. {An expression for the current in terms of $\check{g}%
_{R(L)}$ can be obtained straightforwardly from the equations of motions for
the Green functions after integration over the quasiparticle energy. In the
lowest order in tunneling the current is given by}
\begin{equation}
I=\frac{1}{32eR_{N}(T_{\uparrow }^{2}+T_{\downarrow }^{2})}\int d\epsilon
\mathrm{Tr}\left\{ \hat{\tau}_{3}\left[ \check{\Gamma}\check{g}_{L\alpha }%
\check{\Gamma},\check{g}_{R\beta }\right] ^{K}\right\}   \label{current}
\end{equation}%
where $R_{N}=1/[4\pi (eN(0))^2(T_{\uparrow }^{2}+T_{\downarrow }^{2})]$ is the
resistance of the barrier in the normal state, $N(0)$ is the normal density
of states at the Fermi level, the symbol $\check{.}$ denotes 8$\times $8
matrices in the Gor'kov-Nambu ($\tau _{i}$) - Spin ($\sigma _{i}$) - Keldysh
space, $\check{\Gamma}=\mathcal{T}\hat{\tau}_{0}\otimes\hat{\sigma}_{0}+\mathcal{U}%
\hat{\tau}_{3}\otimes\mathbf{\hat{\sigma}}_{3}$, $\alpha $, $\beta $ are the angles
between the exchange field of the L and R electrode with respect to the
z-axis (see inset in Fig. 1), and $\check{g}_{\alpha }$ is the bulk Green's
function which can be obtained by solving the  quasiclassical
equations.
 The matrix $\check{g}_{L\alpha }$ (and in analogy  $\check{g}_{R\beta }$) can be written as 
$\check{g}_{L\alpha }=\check{R}_{\alpha } .\check{g}_{L0}.\check{R}_{\alpha }^{\dagger }$, where $\check{g}_{L0}$  is   the known solution for the case of an exchange field along the $z-$axis,  and 
$\check{R}_{\alpha }=\cos (\alpha
/2)+i\hat{\tau}_{3}\otimes \hat{\sigma}_{1}\sin (\alpha /2)$  

\textit{Results} We first proceed to determine the Josephson critical
current through the spin-filter. We assume that $\Delta _{L}=\Delta
_{R}=\Delta $ and ${h_{L}}={\ h_{R}}=h$ and that the exchange field in the
left (right) electrode forms an angle $\alpha $ ($\beta $) with the
magnetization of the I$_{sf}$ barrier which points in z-direction. From Eq. (%
\ref{current}) we find the Josephson current $I_J=I_c\sin\varphi$, where $%
\varphi$ is the phase difference between the superconductors and the
critical current $I_{c}$ is given by the general expression:
\begin{equation}
eR_{N}I_{c}=2\pi T\sum_{\omega _{n}>0}\left\{ r\left[ f_{s}^{2}+f_{t}^{2}%
\cos \alpha \cos \beta \right] +f_{t}^{2}\sin \alpha \sin \beta \right\} \;,
\label{jos_curr}
\end{equation}%
where $r=2T_{\downarrow }T_{\uparrow }/(T_{\downarrow }^{2}+T_{\uparrow
}^{2})$ is a parameter describing the efficiency of the spin-filtering ($r=0$
denotes full polarization while $r=1$ a non-magnetic barrier\cite{note}) . $%
f_{s(t)}=(f_{+}\pm f_{-})/2$ are the anomalous Green's functions where $%
f_{\pm }=\Delta /\sqrt{(\omega _{n}\pm ih)^{2}+\Delta ^{2}}$ and $\omega _{n}
$  the Matsubara frequency. The amplitude of the singlet component is
determined by $f_{s}$ whereas the amplitude of the triplet component is
given by $f_{t}$. Eq. (\ref{jos_curr}) is one of the main results of our
work. If $h=0$ it reproduces the expression presented in Refs. \cite%
{Kulik:1966us,bula_kuzi} which is the well-known Ambegaokar-Baratoff (AB)
formula for the critical current\cite{Ambegaokar:1963fk} multiplied by a
factor $r<1$.

In the case of a fully spin-polarized barrier ($r=0$), \textit{i. e.} if
either $T_{\downarrow }$ or $T_{\uparrow }$ is zero, Eq.(\ref{jos_curr})
shows that the singlet Cooper pairs do not contribute to the Josephson
current. The contribution to the current is only due to the second term on
the r.h.s. which is independent of $r$ and proportional to the amplitude of
the triplet component $f_t$. This term does not vanish provided that neither
$\alpha $ nor $ \beta$ are equal to $0$ or $\pi$. This important result
shows that even though in the electrodes only the triplet component with
(locally) zero spin projection exists, the non-collinearity between $h$ and
the  magnetization of the barrier induces a coupling between them and
leads to the creation of a spin polarized supercurrent. {In our model 
 the parameters }$r${\ and }$h$%
{\ are independent. However,   for a ferromagnetic insulator/ superconductor system 
  they might  be related to  each other\cite{Tokuyasu:1988et}.}
\begin{figure}[tb]
\includegraphics[width=\columnwidth]{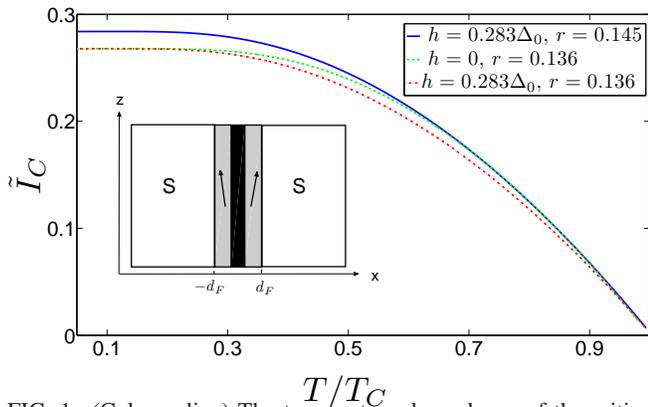} \vspace{-2mm} \vspace{-8mm}.
\caption{(Color online) The temperature dependence of the critical current
for different values of $h$ and $r$. We assume that $\protect\alpha=\protect%
\beta=0$. Inset: The structure described by our model Hamiltonian Eq. (\ref%
{Hamiltonian}). The black region represents the spin-filter barrier while
the grey regions are layers with a finite exchange field pointing in
arbitrary direction. $d_F$ is the thickness of these layers and we have
defined $\tilde I_c=2I_c e R_N/(\Delta_0\protect\pi)$, where $\Delta_0$ is
the value of the order parameter at $T=0$ and $h=0$.
\label{fig1}}
\end{figure}

We assume next that the exchange fields in the left and right electrodes are
parallel to the magnetization of the barrier ($\alpha =\beta =0$) and
compute the temperature dependence of the critical current using Eq. (\ref%
{jos_curr}). In panel Fig.\ref{fig1} we show this dependence for different
sets of parameters $(h,r)$. Throughout this article the order parameter $%
\Delta (T)$ is determined self-consistently and the temperature in the
figures is normalized with respect to the critical temperature which depends
on $h$. The $I_{c}(T)$ curve was measured in Ref. \cite{Senapati:2011fm} for
a Josephson junction with a spin-filter as tunneling barrier. If we assume,
as the authors of Ref. \cite{Senapati:2011fm} did, a finite spin-filtering
effect ($r<1$) but neglect the exchange field in the superconductor ($h=0$)
we obtain the dashed curve in Fig.\ref{fig1}, which is nothing but the AB
curve multiplied by a pre-factor $r\approx 0.27$. If we assume now a finite
value of the effective exchange field in the S layers (dot-dashed curve in
Fig.\ref{fig1}) for the same value of $r$ one obtains that critical current
is for all values of temperatures smaller than the AB curve. If we now keep
the same value for the finite exchange field but slightly change the value
of $r$ ( solid line in Fig.\ref{fig1}), one can see that for lower
temperatures the $I_{c}(T)$ curve exceeds that of the AB curve. This
behavior, which is in qualitative agreement with the results of Ref. \cite%
{Senapati:2011fm}, shows that the interplay between $h$ and $r$ is crucial
to understand the transport properties of the junction. We cannot
conclusively say, though, that the experiment can be fully explained by
these results. Indeed, measurements of the tunneling conductance in
junctions with GdN barriers suggest a finite exchange field inside the S
electrodes\cite{Pal}. However, GdN barriers may also exhibit a complicated
temperature dependent magnetic domain structure that could also modify the $%
I_{c}(T)$ behavior\cite{Senapati:2011fm}. This hypothetical effect is beyond
the scope of the present work.
\begin{figure}[tb]
\includegraphics[width=\columnwidth]{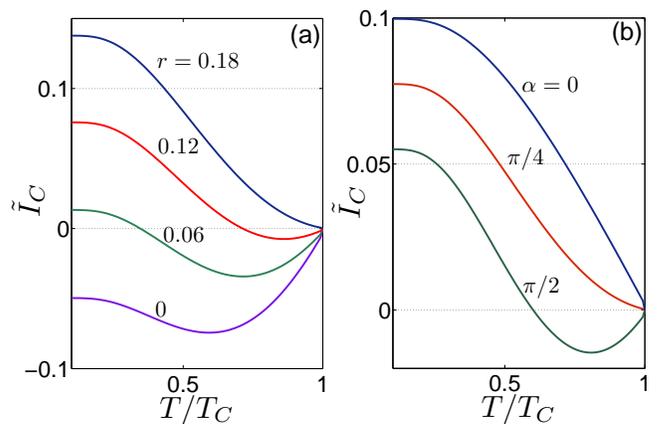} \vspace{-2mm} \vspace{-8mm}.
\caption{(Color online) The temperature dependence of $\tilde{I}_{c}$ for
different values of $r$ and $\protect\alpha =\protect\pi /2$ (a); and for
different values of $\protect\alpha $ and $r=0.1$ (b). In both panels $%
h=0.567\Delta _{0}$
\label{fig2}}
\end{figure}

Let us now assume that the exchange fields in the S layers and the
magnetization of the $I_{sf}$ barrier are non-collinear (we set $%
\alpha=\beta=\pi/2$ in order to maximize the contribution of the triplet
supercurrent). In Fig. \ref{fig2}a we show the temperature dependence of the
critical current for different values of the spin-filter parameter $r$
corresponding to highly polarized barriers. For large values of $r$ the
critical current is positive for all temperatures ($0$-junction). However if
$r$ is small enough the second term in the r.h.s. of (\ref{jos_curr}) start
to dominate and at certain interval of temperature $I_c<0$ ($\pi$-junction),
\textit{i.e.} our model predicts a zero-$\pi$ transition for large enough
spin-filter efficiency. Thus, it is more likely to observe the 0-$\pi$
transition in systems containing europium chalcogenide tunnel barriers with
a almost 100\% spin polarization\cite{Santos:2008cm} than using GdN films
with a spin-filter efficiency of around 75\%\cite{Senapati:2011fm}. Note
that in the fully polarized limit $r=0$ the critical current is negative for
all temperatures. In Fig. \ref{fig2}b we show the $I_c(T)$ dependence for $%
r=0.1$ and different values of $\alpha$. Negative values of the current
appear if $\alpha$ is close to $\pi/2$.  The origin of the $\pi$ junction
behavior described here is different from the one studied in Ref. \cite%
{bula_kuzi}. The $\pi$-junction behavior shown in Fig. 2 is caused by the
non-collinearity of the exchange fields and the magnetization of the $I_{sf}$%
, \textit{i.e.} it is determined by the second term in the r.h.s of Eq. (\ref%
{jos_curr}). In contrast, in Ref. \cite{bula_kuzi} there is no such term and
the $\pi$ junction behavior was obtained by assuming that $\mathcal{T}<%
\mathcal{U}$ (\textit{i.e.} by choosing $r<0$).
{For completeness we note that the Josephson current in metallic 
 multilayered SFFS junctions also depends on the angle of mutual magnetization orientations in different F
layers. This problem (without spin-filter barriers) was studied in numerous papers on the basis of Usadel,
Eilenberger or Bogoliubov-de Gennes equations (see for example \cite%
{Volkov:2003hl,Buzdin2006,Ivanov2007,Braude2007,Houzet:2007ib,Valls2007,Linder2008,Volkov2009,x1,x2}
{and references in the review articles} \onlinecite{Bergeret:2005bf,Eschrig:2011ht}).}

Let us now calculate the quasiparticle current $I_{qp}$ from Eq.(\ref%
{current}). For the normalized current $j_{qp}=I_{qp}(V)/I_{N}(V)$\ ($%
I_{N}(V)=V/R_{N}$\ is the current through the junction in the normal state)
we get
\begin{equation}
j_{\alpha \beta }=\frac{1}{eV}\int d\epsilon F_{V}Y_{\alpha \beta }(\epsilon
,h,V)\; ,  \label{QPcurrent}
\end{equation}
where $Y_{\alpha \beta }(\epsilon ,h,V)$ is the spectral conductance and $%
F_V=0.5\{\tanh[(\epsilon+eV/2)/2T]-\tanh[(\epsilon-eV/2)/2T]\}$. We present
here the expression for a symmetric junction, \textit{i.e. i.e.} $\nu
_{r}=\nu _{l}$ and $\alpha=\beta$, although similar expressions hold for
arbitrary angles $\alpha$ and $\beta$. It reads,
\begin{equation}
Y_{\alpha \alpha }=\nu _{0+}\nu _{0-}+\nu _{3+}\nu _{3-}-(1-r)\nu _{3+}\nu
_{3-}\sin ^{2}\alpha\; .  \label{yaa}
\end{equation}%
We have defined the DOS $\nu _{0,3}(\epsilon )=[\nu (\epsilon +h)\pm \nu
(\epsilon -h)]/2$\ with $\nu (\epsilon )=\epsilon /\sqrt{\epsilon
^{2}-\Delta ^{2}}$ and $\nu _{0\pm }=\nu _{0}(\epsilon \pm eV/2)$ .

The left panel of Fig. 4 shows the voltage dependence of the normalized
differential conductance $G_{qp}$ for zero-temperature in the symmetric case
$\alpha=\beta=\pi/4$. In the  absence of spin-filter effect ($r=1$) the
differential conductance is an even function of $V$, showing a peak at $%
eV=2\Delta$ and no signature of the exchange splitting\cite{Mersevey1994}
(dash-dotted line in the left panel of Fig. 4). However for $r<1$ and $%
0<\alpha<\pi$ two additional peaks appear at $eV=2(\Delta \pm h)$. Notice
that the height of these peaks increases with decreasing $r$. Thus, by
measuring the differential conductance one can extract information about the
model parameters $\alpha$, $r$ and $h$. From Eqs. (\ref{QPcurrent}-\ref{yaa}%
) one can also show that in order to obtain an asymmetric $G_{qp}(V)$
dependence one should set $h_L\neq h_R$ as discussed Ref. \cite{Hao:1990fl}.
In the antiparallel case, where $\alpha=0$, $\beta=\pi$ (right panel of Fig.
4) and for $r<1$ the differential conductance has peaks at $eV=2(\Delta \pm
h)$ (but not at $eV=2\Delta$). These peaks have the same size for $r=1$.
However, by decreasing $r$ towards zero, the difference between the peak
sizes increases. In the fully spin-polarized case ($r=0$) one of these peaks
vanishes.
\begin{figure}[tb]
\includegraphics[width=\columnwidth]{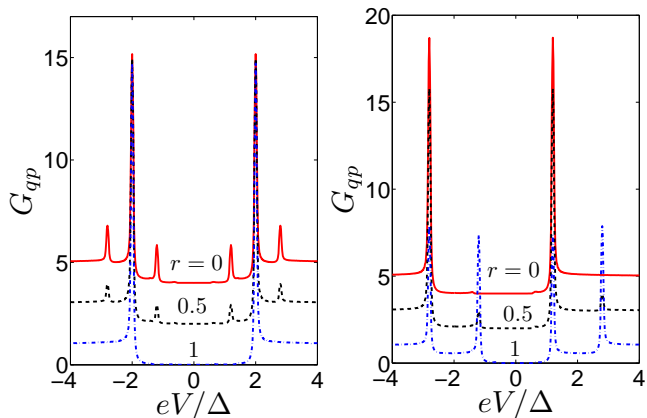} \vspace{-1mm} \vspace{-8mm}.
  \caption{(Color online) The zero-temperature normalized differential
conductance $G_{qp}={R_N dI_{qp}}/{dV}$ for $h=0.4\Delta_0$ and $r=0,0.5,1$.
Left panel: $\protect\alpha=\protect\beta=\protect\pi/4$, right panel: $%
\protect\alpha=0$, $\protect\beta=\protect\pi$. By calculating the curves we
have added a small $\protect\eta=0.01\Delta_0$ damping factor. Note that the
curves are shifted vertically for clarity}
\label{Fig4}
\end{figure}

\textit{Discussions and Conclusions} Our model can describe different
systems. First, the model applies to junctions made of two magnetic
superconductors separated by a spin-filter barrier $I_{\mathrm{sf}}$. But it
can also describe a S-F-I$_{\mathrm{sf}}$-F-S junction with the width of the
F-S electrodes smaller than the characteristic length over which the Green
functions vary. We have also verified that our results are qualitatively
valid for long S electrodes. These results will be discussed in more detail
elsewhere.  Finally, our model can also describe a simple S-I$_{\mathrm{sf}}$%
-S assuming that the I$_{sf}$ barrier induces an effective exchange field in
the superconductor over distances of the order of the superconducting
coherence length as predicted in Ref. \cite{Tokuyasu:1988et}.

In conclusion, by combining the quasiclassical Green functions and the
tunneling Hamiltonian approach we have studied the effect of {spin filtering
on the Josephson and quasiparticle current in tunneling junctions}. We have
shown that for fully polarized barriers the singlet component does not
contribute to the supercurrent $I_{J}$. However, if the direction of the
exchange field $\mathbf{h}$ in both electrodes is not parallel to the
quantization axis of the barrier, a non-zero $I_{J}$ current is observed due
to the triplet component. In this case the current is 100\% spin polarized.
We have also calculated the differential conductance and shown its
dependence on the spin-filter parameter $r$ and the misalignment angle. By
measuring the differential conductance one can extract information about the
magnetic structure of the spin-filter junction.  Our findings are relevant
for the creation, control and manipulation of spin polarized supercurrents
as well as for the characterization of S-I$_{sf}$-S junctions.

\textit{Acknowledgements} We thank Mark Blamire, Jason Robinson, Avradeep
Pal, Ilya Tokatly and Alberto Rojo for useful discussions and David Pickup
for careful reading of our manuscript. The work of F.S.B and A. V. was
supported by the Spanish Ministry of Economy and Competitiveness under
Project FIS2011-28851-C02-02 and the Basque Government under UPV/EHU Project
IT-366- 07. A. F. V. is grateful to the DIPC for hospitality and financial
support.
%


\begin{thebibliography}{10}
\bibitem{Footnote} {If the   magnetization inhomogeinity  
occurs on a scale of the order of the Fermi wave length, one can talk 
about  a  spin-active interface \cite{Eschrig03}.}
\bibitem{note} Throughout this paper we take values for $r$ such that $0<r<1$ ({\it i.e.} for ${\cal T}\geq {\cal U}$). However, all the expressions in the present work   are also valid for the case ${\cal T}<{\cal U}$.{ The spin-filter efficiency is usually defined as\cite{Senapati:2011fm}   $P=|T_\uparrow^2-T_\downarrow^2|/(T_\uparrow^2+T_\downarrow^2)$, thus the relation between the parameter $r$ and $P$ is given by $P^2=1-r^2$}.
\end{thebibliography}

\end{document}